\documentclass[aps,prl,twocolumn,reprint,amsmath,amssymb,superscriptaddress,floatfix,footinbib,longbibliography]{revtex4-1}

\usepackage{epsfig}
\usepackage{graphicx}
\usepackage{epstopdf}

\usepackage[T1]{fontenc}
\usepackage[applemac]{inputenc}
\usepackage{lmodern}
\usepackage[english]{babel}

\usepackage{ae}
\usepackage{units}

\usepackage{amsmath,amssymb,natbib,bm}
\usepackage{psfrag}
\usepackage{subfigure}
\usepackage{amsthm}
\usepackage{bbm}

\usepackage{booktabs}

\usepackage[americaninductors]{circuitikz}
\usepackage{tikz}
\usetikzlibrary{arrows}

\usepackage{color}
\usepackage{url}

\usepackage[colorlinks]{hyperref}
\hypersetup{%
        plainpages=true,
        breaklinks=true,
        hypertexnames=false,
        pageanchor=true,
        colorlinks=true,
        linkcolor={blue},
        citecolor={magenta},
        urlcolor={blue},
        anchorcolor={black}
      }
      
\usepackage{mleftright} 

\newcommand{\figref}[1]{\mbox{Fig.~\ref{#1}}}

\renewcommand{\eqref}[1]{\mbox{Eq.~(\ref{#1})}}

\newcommand{\ket}[1]{\mleft|#1 \mright \rangle}

\newcommand{\ketbra}[2]{\mleft| #1 \rangle \langle #2 \mright|}

\newcommand{\tr}[1]{\text{tr}\mleft( #1 \mright)}

\newcommand{\abs}[1]{\mleft|#1\mright|}

\newcommand{\be}{\begin{equation}}
\newcommand{\ee}{\end{equation}}
\newcommand{\bea}{\begin{eqnarray}}
\newcommand{\eea}{\end{eqnarray}}

\AtBeginDocument{%
    \newwrite\bibnotes
    \def\bibnotesext{Notes.bib}
    \immediate\openout\bibnotes=\jobname\bibnotesext
    \immediate\write\bibnotes{@CONTROL{REVTEX41Control}}
    \immediate\write\bibnotes{@CONTROL{%
    apsrev41Control,author="08",editor="1",pages="0",title="0",year="1"}}
     \if@filesw
     \immediate\write\@auxout{\string\citation{apsrev41Control}}%
    \fi
}%

\begin{document}

\title{Quantum State Tomography with Conditional Generative Adversarial Networks}
\date{\today}

\author{Shahnawaz Ahmed}
\email{shahnawaz.ahmed95@gmail.com}
\affiliation{Department of Microtechnology and Nanoscience, Chalmers University of Technology, 412 96 Gothenburg, Sweden}

\author{Carlos S\'anchez Mu\~noz}
\affiliation{Departamento de Fisica Teorica de la Materia Condensada and Condensed Matter Physics Center (IFIMAC), Universidad Autonoma de Madrid, Madrid, Spain}

\author{Franco Nori}
\affiliation{Theoretical Quantum Physics Laboratory, RIKEN Cluster for Pioneering Research, Wako-shi, Saitama 351-0198, Japan}
\affiliation{Department of Physics, University of Michigan, Ann Arbor, Michigan 48109-1040, USA}

\author{Anton Frisk Kockum}
\email{anton.frisk.kockum@chalmers.se}
\affiliation{Department of Microtechnology and Nanoscience, Chalmers University of Technology, 412 96 Gothenburg, Sweden}

\begin{abstract}
Quantum state tomography (QST) is a challenging task in intermediate-scale quantum devices. Here, we apply conditional generative adversarial networks (CGANs) to QST. In the CGAN framework, two duelling neural networks, a generator and a discriminator, learn multi-modal models from data. We augment a CGAN with custom neural-network layers that enable conversion of output from any standard neural network into a physical density matrix. To reconstruct the density matrix, the generator and discriminator networks train each other on data using standard gradient-based methods. We demonstrate that our QST-CGAN reconstructs optical quantum states with high fidelity orders of magnitude faster, and from less data, than a standard maximum-likelihood method. We also show that the QST-CGAN can reconstruct a quantum state in a single evaluation of the generator network if it has been pre-trained on similar quantum states.
\end{abstract}

\maketitle


\paragraph*{Introduction.}

The ability to manipulate and control small quantum systems opens up promising directions for research and technological applications: quantum information processing and computation~\cite{Feynman1982, Montanaro2016, Wendin2017, Preskill2018, Arute2019}, simulations of quantum chemistry~\cite{Georgescu2014, Kandala2017, Childs2018, Arguello-Luengo2019, Ma2020}, secure communication~\cite{Kimble2008, Yin2020}, and much more~\cite{You2011, Georgescu2014, Gu2017, Bal2012, Degen2017, Pirandola2018, Gefen2019, Wang2020, Lloyd2008, Korobko2019, Barzanjeh2020}. A prominent example is the recent demonstration of a 53-qubit quantum computer performing a computational task in a few hundred seconds that was anticipated to take much longer on a classical supercomputer~\cite{Arute2019}. Such speedup is possible partly due to the exponentially large state space that can be used for storage and manipulation of information in quantum systems~\cite{Nielsen2000, Caves2004, Havlicek2019}. However, this large size of the state space also brings challenges for the characterization and description of these systems.

The process of reconstructing a full description of a quantum state by measuring its properties is called quantum state tomography (QST)~\cite{DAriano2003, Liu2005, Lvovsky2009}. Tomography is fundamentally a data processing problem, trying to extract meaningful information from as few (noisy) measurements as possible~\cite{Deleglise2008, DAriano2009, Cramer2010, Flammia2011, Petz2012, Baumgratz2013, Miranowicz2015, Hou2016, Titchener2018, Rocchetto2019, Huang2020}. There exist many cleverly crafted QST techniques apart from general maximum likelihood estimation (MLE)~\cite{Banaszek2000, Lvovsky2004}, e.g., diluted MLE~\cite{Rehacek2007}, compressed sensing~\cite{Gross2010}, Bayesian tomography~\cite{Blume-Kohout2010, Granade2016}, projected gradient descent~\cite{Bolduc2017}, matrix-product-state and tensor-network tomography~\cite{Cramer2010, Lanyon2017, Glasser2018}, and permutationally invariant tomography~\cite{Toth2010, Moroder2012}. However, these techniques are often restricted to specific types of quantum states, lacking versatility~\cite{Carrasquilla2019}.

Recently, machine-learning methods have been applied to QST, yielding promising results~\cite{Carleo2017, Xu2018, Torlai2018, Torlai2018a, Xin2018, Palmieri2020, Gray2018, Quek2018, Yu2019, Melkani2019, Weiss2019, Liu2020, Xin2020}. In particular, generative models~\cite{Lloyd2018, Kieferova2017, Carrasquilla2019, Hu2019}, usually restricted Boltzmann machines (RBMs), have been used as Ans\"{a}tze with few parameters to represent a quantum state and learn the probability distribution of outputs expected from that state~\cite{Carleo2017, Glasser2018, Tiunov2020}. There are also examples of deep neural networks being used for QST~\cite{Carleo2018, Rocchetto2018, Cai2018, Palmieri2020, Cha2020, Lohani2020}, enabling physicists to take advantage of the rapid progress in such machine-learning techniques.

One interesting recent development in machine learning is generative adversarial networks (GANs)~\cite{Goodfellow2014, Mahdizadehaghdam2019}. Such networks have led to an explosion of new results that were previously thought futuristic: generation of photorealistic images~\cite{Isola2017, Karras2019, Karras2019a}, conversion of sketches to images~\cite{Isola2017}, text generation in different styles~\cite{Yang2018, Subramanian2018}, text-to-image generation~\cite{Xu2018a}, generating and defending against fake news~\cite{Mirsky2020, Zellers2019}, and even game design learned from observing video~\cite{Kim2020}. An improvement of standard GANs that led to many of these results is conditional generative adversarial learning~\cite{Mirza2014}, which enabled increased control of the output of generative models. Recently, such GANs have been applied to tomography of materials structure with synchrotron radiation~\cite{Yang2020, Liu2020a} and computed tomography of soft tissue in medicine~\cite{Selim2020}.

In this Letter, we introduce QST with conditional GANs (QST-CGAN). Leveraging a CGAN architecture, complemented by custom layers for representing a quantum state in the form of a density matrix, we show that adversarial learning can be a powerful tool for QST. The QST-CGAN is different from RBM-based methods since it learns a map between the data and the quantum state instead of a probability distribution. The custom layers we introduce bridge a gap between ML and quantum information processing; they enable many further applications beyond the QST-CGAN presented here. We benchmark the QST-CGAN on reconstruction of various simulated optical quantum states, and show an example with real experimental data. The QST-CGAN performance is superior to that of a standard maximum-likelihood reconstruction method in terms of reconstruction fidelity, convergence time, and amount of measurement data required. We also show that a QST-CGAN can reconstruct quantum states in a single pass through the network if it has been pre-trained on simulated data.

Our reconstruction method is versatile, general, and ready to be applied for QST of intermediate-scale quantum systems, which are widely explored in current experiments~\cite{Preskill2018}. In Refs.~\cite{Ahmed2020, QST-CGAN2020}, we provide more details on our implementation (including data and code) and also discuss classification of quantum states with neural networks.


\paragraph*{Quantum state tomography with maximum likelihood estimation.}

Quantum state tomography estimates the quantum state (a state vector $\ket \psi$ or a density matrix $\rho$) from measurements of Hermitian operators $\mathcal O$~\cite{Raymer1997, Lvovsky2009}. The operators are usually positive-operator-valued measures (POVMs), a set of positive semi-definite matrices $\{\mathcal O_i\}$ that sum to identity, $\sum_i^k \mathcal O_i = I$, representing a measurement with $k$ possible outcomes. The probability of each outcome is given by $\tr{\mathcal O_i \rho}$. A set of operators that allows for the complete characterization of a quantum state is called informationally complete (IC)~\cite{DAriano2004}.

In an experiment, single-shot measurements are repeated over an ensemble of identical states to collect statistics: the frequencies $d_i$ of POVM outcomes. These frequencies give an estimate of the expectation values $\tr{\mathcal O_i \rho}$, where $\rho$ is the density matrix describing the state. The outcomes of many different POVMs can be combined to form a linear system of equations $\mathbf d = A \rho_f$, where $\rho_f$ is the flattened density matrix and $A$ is the ``sensing matrix'' determined by the choice of POVMs~\cite{Shen2016}. Solving this system of equations by linear inversion methods to obtain $\rho$ can fail, either due to the statistical nature of the (noisy) measurement or due to a high condition number for $A$~\cite{Shen2016}.

An alternative to linear inversion methods is maximum likelihood estimation (MLE). In MLE, the likelihood function~\cite{Banaszek2000, Shang2017} $L (\rho^{\prime} | \mathbf d) = \prod_{i} [\tr{\rho^{\prime} \mathcal O_i}]^{d_i}$ is maximized to find the best estimate $\rho^{\prime}$ for reproducing the experimental data. In this Letter, we take a different approach by applying CGANs to find $\rho'$.


\paragraph*{Conditional generative adversarial networks.}

In generative adversarial learning, a generator $G$ and a discriminator $D$ compete to learn a mapping from some prior noise distribution to a data distribution~\cite{Goodfellow2014}. The generator and the discriminator are parameterized nonlinear functions [parameters $(\mathbf \theta_D, \mathbf \theta_G)$], usually multi-layered neural networks. The generator takes an input $\mathbf z \sim p_z (\mathbf z)$ from the noise distribution $p_z (\mathbf z)$ and generates an output $G(\mathbf z; \mathbf \theta_G)$. The discriminator takes an input $\mathbf q$ and outputs a probability $D(\mathbf q; \mathbf \theta_D)$ that it belongs to the data distribution $p_{\texttt{data}}$.

The parameters of G and D are optimized alternatively such that the generator produces outputs that resemble the data and thus fool the discriminator, and the discriminator becomes better at detecting fake (generated) output. In each optimization step, $\mathbf \theta_D$ is updated to maximize the expectation value
\be
E_{\mathbf y \sim p_{\texttt{data}}} [ \log(D(\mathbf y; \mathbf \theta_D))] + E_{\mathbf z \sim p_z} [\log (1 - D(G(\mathbf z; \mathbf \theta_G); \mathbf \theta_D))],
\label{eq:MaximizeD}
\ee
where $\mathbf{y}$ denotes samples from the data. Then, $\mathbf \theta_G$ is updated to minimize
\be
E_{\mathbf z \sim p_z} [\log (1 - D(G(\mathbf z; \mathbf \theta_G); \mathbf \theta_D))].
\label{eq:MinimizeG}
\ee
In this way, the generator learns to map elements from a noise distribution to data as $G: \mathbf z \to \mathbf y$~\cite{Goodfellow2014, Isola2017}. 

However, since the generator input is random, we have no control over the output. This issue is solved by using a conditional generative adversarial network (CGAN)~\cite{Mirza2014, Isola2017}. In a CGAN, the generator and discriminator output is conditioned on some variable $\mathbf x$. This conditioning allows the generator to learn the mapping $G: {\mathbf x, \mathbf z} \to \mathbf y$~\cite{Isola2017}. The optimization of parameters for the CGAN is done as before, by maximizing \eqref{eq:MaximizeD} and minimizing \eqref{eq:MinimizeG}; the only difference is that the outputs now are $D(\mathbf {x, y};\theta_D)$ and $G(\mathbf {x, z};\theta_G)$. This CGAN approach is very flexible and can be used to find complex maps between inputs and outputs. The flexibility stems from using the discriminator network for evaluation instead of, or in addition to, a simpler loss function.


\begin{figure}
\centering
\includegraphics[width=\linewidth]{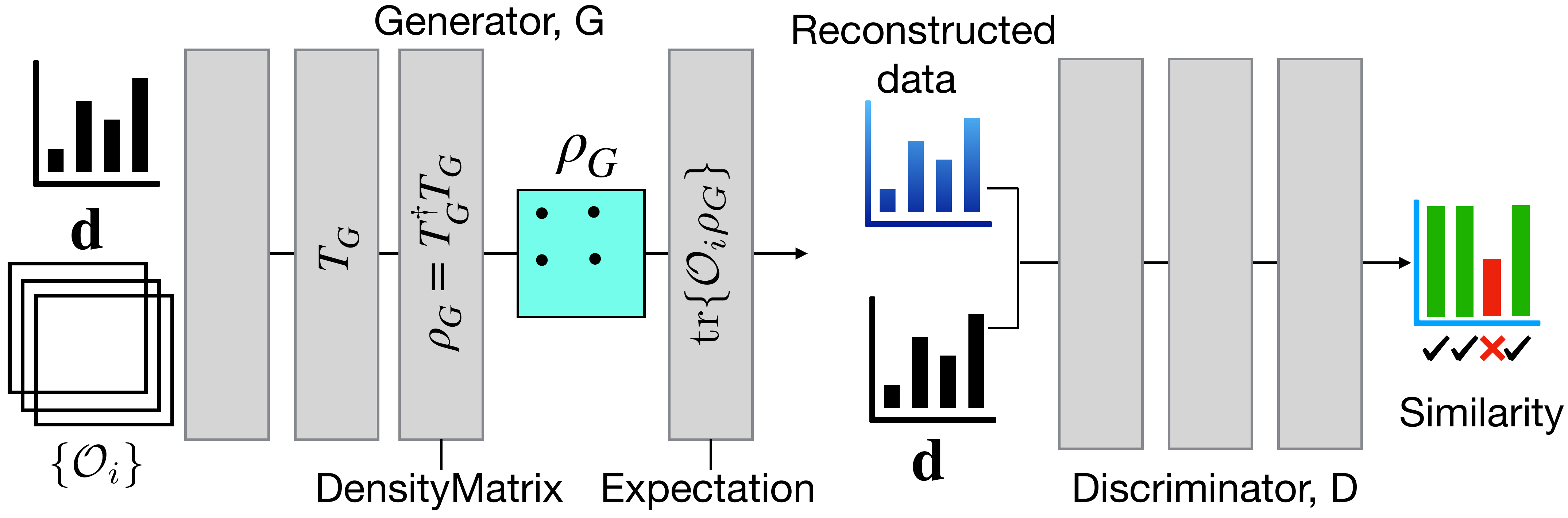}
\caption{Illustration of the CGAN architecture for QST. Data $\mathbf d$ sampled from measurements of a set of measurement operators $\{\mathcal O_i \}$ on a quantum state is fed into both the generator $G$ and the discriminator $D$. The other input to $D$ is the generated statistics from $G$. The next to last layer of $G$ outputs a physical density matrix and the last layer computes measurement statistics using this density matrix. The discriminator compares the measurement data and the generated data for each measurement operator and outputs a probability that they match.}
\label{fig:fake}
\end{figure}

\paragraph*{Quantum state tomography using conditional generative adversarial networks.}

We now adapt the CGAN framework to the problem of QST. In our approach, illustrated in \figref{fig:fake}, the conditioning input to the generator is the measurement statistics and the measurement operators ($\mathbf x \to \textbf d, \{\mathcal O_i \}$). The generated output is a density matrix $\rho_G$. We find that we do not need to provide any input noise $\mathbf z$, consistent with the results in Ref.~\cite{Isola2017}.

The discriminator takes as input the experimental measurement statistics $\textbf d$ (as the conditioning variable) and generated measurement statistics calculated from $\tr {\mathcal O_i \rho_{G}}$. The output from the discriminator is a set of numbers describing how well the generated measurement statistics match the data. This partitioning of the evaluation of the generated statistics is inspired by the \textit{Patch}GAN architecture of Ref.~\cite{Isola2017}. If the generator has managed to learn the correct density matrix, the discriminator will not be able to distinguish the generated statistics from the true data.

The adaption of the CGAN architecture to QST requires us to introduce two custom layers at the end of the generator neural network. First, we add a \textit{DensityMatrix} layer, which takes the unconstrained intermediate output of the generator, moulds it into a lower triangular complex-valued matrix $T_G$ with real entries on the diagonal, constructs $T_G^\dag T_G$, and normalizes the resulting matrix to have unit trace. This method is inspired by the Cholesky decomposition~\cite{Banaszek2000}. It ensures that the output $\rho_G$ is a valid density matrix: Hermitian, positive, and having unit trace. A similar idea was found independently in Ref.~\cite{Lohani2020}.

Secondly, we add an \textit{Expectation} layer that combines the output $\rho_G$ with the given measurement operators $\{\mathcal O_i \}$ to compute the generated measurement statistics for each measurement outcome as $\tr{\mathcal O_i \rho_G}$. These two custom layers do not have any trainable parameters. They are only present to enforce the rules of quantum mechanics in the neural networks. This is akin to regularization~\cite{Hinton2012} and normalization~\cite{Hoffer2018} in neural networks. We note that our two custom layers could be used to augment any deep-learning neural-network architecture for QST, e.g., Refs.~\cite{Cha2020, Lohani2020}.

We train the QST-CGAN using standard gradient-based optimization techniques, e.g., Adam~\cite{Kingma2015} with learning-rate scheduling, starting from random initial values for the parameters $(\mathbf \theta_D, \mathbf \theta_G)$. In this way, data from one experiment can be used  to estimate the density matrix of the state in that experiment. However, when reconstructing $\rho$ from another experiment, the QST-CGAN must start from zero again. We can avoid this reset by pre-training on simulated data corresponding to the type of state(s) and noise that is expected to be present in the experiment. The reconstruction from experimental data then requires less additional training; it even becomes possible to do \textit{single-shot reconstruction} with a single evaluation by the pre-trained generator.

We note that adding $L_1$ loss to \eqref{eq:MinimizeG} as suggested in Ref.~\cite{Isola2017} proved helpful in training the QST-CGAN~\cite{Ahmed2020}, and was used for all results displayed below, but was not necessary to obtain good results. Similarly, adding a gradient penalty~\cite{Gulrajani2017} to \eqref{eq:MaximizeD} improved results for single-shot reconstruction.


\paragraph*{Benchmarking CGAN quantum state tomography.}

To benchmark the QST-CGAN method, we test it on reconstruction of optical quantum states and compare its performance to a standard MLE method: iterative MLE (iMLE)~\cite{Lvovsky2004}. In iMLE, projection operators determined by the measurement statistics are iteratively applied to a random initial density matrix until convergence. The final result is an estimated density matrix $\rho'$ that maximizes the likelihood function $L (\rho^{\prime}|\mathbf{d})$.

Optical quantum states describe quantized single-mode electromagnetic fields (harmonic oscillators). Our choice of optical quantum states for testing the QST-CGAN was motivated by the existence of visual representations, e.g., Wigner functions, for these states, seeing how CGANs have mainly been applied to image processing. However, we stress that the QST-CGAN approach is general and can be applied to any type of quantum system with any type of observable~\cite{Ahmed2020}.

\begin{figure}
\centering
\includegraphics[width=\linewidth]{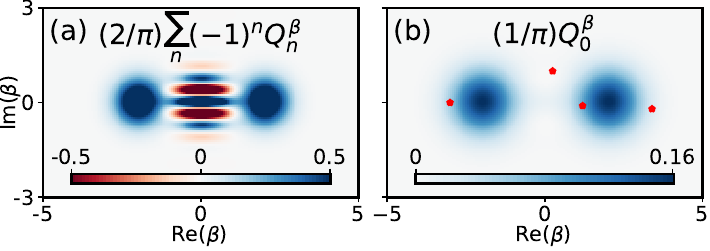}
\caption{Observables for an optical quantum state, the ``cat state'' $\ket{\alpha} + \ket{-\alpha}$ (up to a normalization), with coherent amplitude $\alpha = 2$. (a) The Wigner function. (b) The Husimi Q function. The stars mark specific $\beta$ used to sample the data used as input to the QST-CGAN.}
\label{fig:data}
\end{figure}

Some of the common observables for optical quantum states are instances of a displace-and-measure technique. For example, the photon-number distribution obtained after applying a displacement $\beta$ is the generalized $Q$ function~\cite{Kirchmair2013}: $Q_n^{\beta} = \tr{\ketbra{n}{n} D(-\beta) \rho D^{\dagger}(-\beta)}$, where $\ket{n}$ is the Fock state with $n$ photons, $D(\beta) = e^{\beta a^\dagger - \beta^* a}$ is the displacement operator, and $a (a^{\dagger})$ is the bosonic creation (annihilation) operator of the electromagnetic mode. The Husimi $Q$ function (photon field quadratures) is $\mleft( 1 / \pi \mright) Q^{\beta}_0$ and the Wigner function (photon parity) is $W(\beta) = \mleft( 2 / \pi \mright) \sum_n (-1)^n Q_n^{\beta}$. The measurement data we consider in the following are samples of $Q^{\beta}_0$ and $W(\beta)$ at certain $\beta$, as illustrated in \figref{fig:data}.

A state $\rho$ in a truncated Hilbert space with size $N$ is specified by up to $N^2 - 1$ real numbers~\cite{Sych2012, Shen2016} (we use $N = 32$). Thus, in general, IC requires displacements and measurements to be carried out such that $\mathbf d$ has at least $N^2 - 1$ elements. However, note that the required number of elements in $\mathbf d$ for reconstruction can be lower, $\propto r N$, if $\rho$ has low rank $r$~\cite{Kalev2015}.


\paragraph*{Results.}

\begin{figure}
\centering
\includegraphics[width=\linewidth]{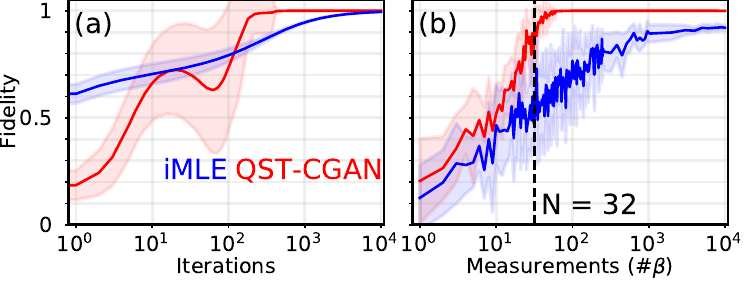}
\caption{QST-CGAN performance. The data is the Husimi $Q$ function of the cat state in \figref{fig:data}(b). (a) Reconstruction fidelity $F (\rho, \rho') = \tr{\sqrt{\sqrt{\rho} \rho' \sqrt{\rho}}}^2$ as a function of iterations for the QST-CGAN (red) and the iMLE (blue).  We use 1024 displacements $\beta$ in a $32 \times 32$ grid. In each of a total of 100 runs, the weights of the QST-CGAN and the starting density matrix of the iMLE are randomly initialized. The solid lines show the mean $F$; the shaded areas show one standard deviation from the mean. (b) Average $F$ as a function of the number of $\beta$. For each number, 10 sets of displacements are randomly selected from within a disk with $\abs{\beta} \le 5$ for the state in \figref{fig:data}. We show the average $F$ reached after 1000 iterations only.
\label{fig:performance}}
\end{figure}

In \figref{fig:performance}(a), we compare the reconstruction fidelity for the QST-CGAN and iMLE methods as a function of the number of iterations. One iteration is one update of all the weights $(\mathbf \theta_D, \mathbf \theta_G)$ for the QST-CGAN (a single gradient-descent step) and one application of the projection operators in iMLE. We find that the QST-CGAN converges to a fidelity $> 0.999$ in about two orders of magnitude fewer iterations (approximately one order of magnitude less time) than the iMLE. Note that the choice of network architecture and training parameters will affect the speed of convergence and the computational cost of one iteration for the QST-CGAN.

Next, we investigate, in \figref{fig:performance}(b), how many data points are required as input to reach high reconstruction fidelity. We find that the QST-CGAN approach starts outperforming the iMLE around $N = 32$ data points and reaches fidelities close to unity already with $<100$ data points, while the iMLE requires $\sim 1000$ data points to attain good fidelity (an RBM-based reconstruction of a similar state also requires thousands of data points to reach high fidelity~\cite{Tiunov2020}). Note that the rank $r = 1$, since $\rho$ is a pure state.


\paragraph*{Experimental state reconstruction from parity measurements.}

\begin{figure}
\centering
\includegraphics[width=\linewidth]{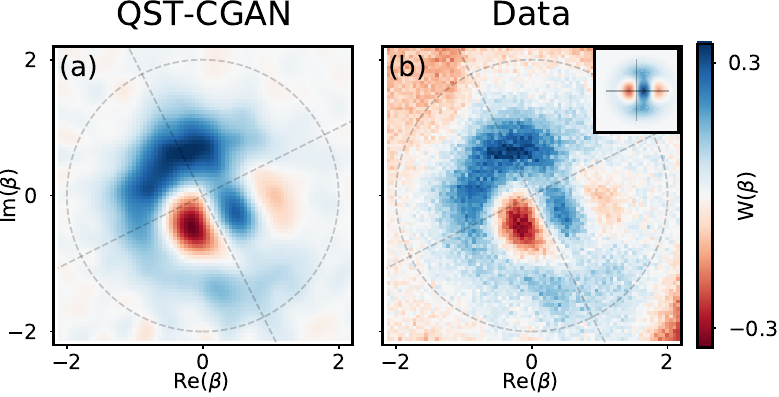}
\caption{(a) Reconstruction of a Wigner-negative state by a QST-CGAN from (b) noisy experimental data. Inset: the target state. The reconstruction uses $4281$ data points of the Wigner function measured for $\beta$ inside the dashed circle. The data outside the circle, e.g., the Wigner-negative region in the top left, is not as reliable due to measurement calibration problems at higher photon numbers. We also attempt reconstruction with a subset of the data points inside the circle, and find that $\sim 600$ data points are enough to achieve a fidelity $\sim 0.9$ with the full reconstruction.}
\label{fig:experiment}
\end{figure}

The benchmarking of the QST-CGAN so far has been on simulated data. We now demonstrate, in \figref{fig:experiment}, that our QST-CGAN can reconstruct a noisy state from experimental data. In this particular experiment, a superconducting transmon qubit was used to generate a Wigner-negative state in a resonator~\cite{Kudra2020}, by applying a selective number-dependent arbitrary phase (SNAP)~\cite{Heeres2015, Fosel2020} of $\pi$ to $\ket{0}$ and $\ket{1}$ of a coherent state $\ket{\alpha = 1}$. Despite significant state-preparation-and-measurement (SPAM) noise, the QST-CGAN still manages to reconstruct the data well from measurements of the Wigner function, even when using only $\sim \unit[15]{\%}$ of the measurement data.


\paragraph*{Single-shot reconstruction with pre-training.}

\begin{figure}
\centering
\includegraphics[width=\linewidth]{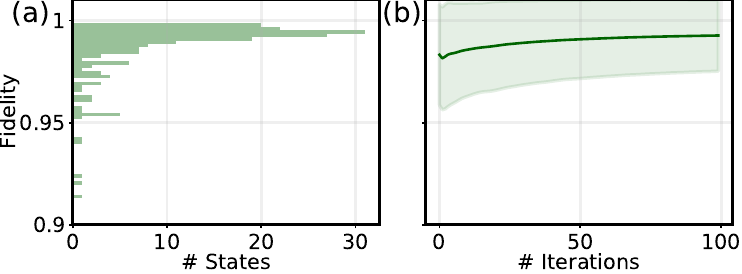}
\caption{Single-shot reconstructions of 200 cat states (cf.~\figref{fig:data}, $|\alpha| \in [1, 3]$, up to six coherent states in superposition), using a pre-trained QST-CGAN. (a) Fidelity distribution of the reconstructions after training on a $32 \times 32$ grid of data points. (b) Average fidelity (solid line) within one standard deviation (shaded region) after further iterations.
\label{fig:singleshot}}
\end{figure}

We now pre-train the QST-CGAN on a data set with several thousand cat states similar to \figref{fig:data} by selecting $|\alpha| \in [1, 3]$ randomly with up to six coherent states in superposition. As shown in \figref{fig:singleshot}(a), this pre-trained network is then able to perform single-shot reconstructions for different cat states with a high average fidelity $\sim 0.98$. It turned out to be difficult to find a learning strategy enabling further improvement of the fidelity with just a few more iterations for each state, but with tens of iterations a clear improvement is observed [\figref{fig:data}(b)]. The pre-trained network thus does not have to iterate many times from an initial random guess for each state, as is the case for the results in \figref{fig:performance} and most other reconstruction methods in use today, resulting in a four orders of magnitude faster reconstruction than in \figref{fig:performance}(a).


\paragraph*{Conclusion and outlook.}

In this Letter, we have adapted the CGAN architecture for use in quantum state tomography. The adaption relies on the introduction of two custom layers, which enforce the properties of a density matrix and allows calculation of expectation values of measurements. We showed that our QST-CGAN clearly outperforms the standard reconstruction method iMLE: the QST-CGAN consistently reconstructs states with higher fidelity, needing $\sim 100 \times$ fewer iterations and $\sim 10 \times$ fewer data points to do so in the examples we showed. Furthermore, we showed that we can pre-train the QST-CGAN on classes of quantum states and achieve high fidelity for single-shot reconstruction.

Looking to the future, we note that the custom layers we introduced could be included into other types of neural networks, e.g., Transformers~\cite{Cha2020}, for both QST and other applications in quantum information processing. The CGAN approach has potential for denoising measurement data by pre-training on simulated noisy data. We further envisage the application of QST-CGAN for adaptive tomography~\cite{Quek2018, DAriano2009}, by choosing next measurements around the points where the discriminator finds that the reconstructed data does not match the experimental data well~\cite{Hu2019}.


\begin{acknowledgments}

\paragraph*{Acknowledgements.}

The QST-CGAN was written in Tensorflow~\cite{tensorflow2015-whitepaper}. Visualization and generation of quantum states was done in QuTiP~\cite{Johansson2012, Johansson2013}. 
We thank Marina Kudra for providing the experimental data for \figref{fig:experiment}.

We acknowledge useful discussions with Yong Lu, Marina Kudra, Florian Marquardt, Steven Girvin, Simone Gasparinetti, Fernando Quijandr\'ia, and Ingrid Strandberg. We also thank Abhijeet Melkani and Clemens Gneiting for comments on the manuscript.

S.A.~and A.F.K.~acknowledge support from the Knut and Alice Wallenberg Foundation through the Wallenberg Centre for Quantum Technology (WACQT). 

C.S.M acknowledges that the project that gave rise to these results received the support of a fellowship from ``la Caixa'' Foundation (ID 100010434) and from the European Union's Horizon 2020 Research and Innovation Programme under the Marie Sk\l{}odowska-Curie Grant Agreement No.~847648. The fellowship code is LCF/BQ/PI20/11760026.

F.N. is supported in part by: NTT Research, Army Research Office (ARO) (Grant No.~W911NF-18-1-0358), Japan Science and Technology Agency (JST) (via the Q-LEAP program and CREST Grant No.~JPMJCR1676), Japan Society for the Promotion of Science (JSPS) (via the KAKENHI Grant No.~JP20H00134 and the JSPS-RFBR Grant No.~JPJSBP120194828), the Asian Office of Aerospace Research and Development (AOARD), and the Foundational Questions Institute Fund (FQXi) via Grant No.~FQXi-IAF19-06.

\end{acknowledgments}

\bibliography{references2}

\end{document}